\documentstyle[12pt,preprint,aps]{revtex}
\def\wp{\relax}
\def\theequation{\arabic{section}.\arabic{equation}}
\begin{document}
\def\bi{\begin{list}{$\bullet$}{\parsep=0.5\baselineskip
\topsep=\parsep \itemsep=0pt}}
\def\ei{\end{list}}
\def\vm{v_{max}}

\def\phi{\varphi}
\def\-{{\bf --}}
\newcommand{\erf}{{\rm erf}}
\newcommand{\m}{{\cal M}}
\newcommand{\lc}{{\cal l}}
\newcommand{\p}{{\cal P}}
\newcommand{\f}{{\cal F}}
\newcommand{\kk}{{\cal K}}
\newcommand{\gc}{{\cal G}}
\newcommand{\zz}{{\cal Z}}
\newcommand{\ww}{{\bf W}}
\newcommand{\s}{\sigma}
\newcommand{\sv}{\underline{\sigma}}
\newcommand{\om}{\omega}
\newcommand{\la}{\lambda}
\newcommand{\eps}{\varepsilon}
\newcommand{\al}{\alpha}
\newcommand{\pf}{\rightarrow}
\newcommand{\jj}{J_{ij}}
\newcommand{\eck}[1]{\left\langle #1 \right\rangle}
\newcommand{\deck}[1]{\langle\!\langle #1 \rangle\!\rangle}
\newcommand{\un}[1]{\underline{ #1 }}

\newcommand{\eq}[1]{(\ref{#1})}
\newcommand{\formel}[1]{\begin{equation} #1 \end{equation}}
\newcommand{\formlab}[2]{\begin{equation}\label{#1} #2 \end{equation}}
\newcommand{\formarr}[1]{\begin{eqnarray} #1 \end{eqnarray}}
\newcommand{\dds}[1]{\frac{\delta}{\delta i\hat\s(#1)}}
\newcommand{\ddh}[1]{\frac{\delta}{\delta i\hat h(#1)}}

\newcommand{\spur}[1]{{\bf Tr}_{ #1 }}
\newcommand{\tr}[1]{{\bf Tr}_{ #1 }}
\newcommand{\gau}[1]{\deck{ #1 }_z}
\newcommand{\gauss}{\int\limits_{-\infty}^{+\infty}
                  \frac{dz}{\sqrt{2\pi}}e^{-z^2/2}\,}
\newcommand{\cc}{{\hbox to 8pt {\hfill\vrule height 6.5pt \kern-2.6pt {\rm C}
                  \hfill}}}
\renewcommand{\thefootnote}{\fnsymbol{footnote}}
\def\ms#1{\marginpar{$\longleftarrow$ MS?}\mark{#1}}


\title{Discrete stochastic models for traffic flow}

\author{M.\ Schreckenberg$^1$,
	A.\ Schadschneider$^1$,
	K.\ Nagel$^2$,
	and N.\ Ito$^{3,}$\footnote{Present address: Dep.~of Applied Physics,
        Faculty of Engineering, University of Tokyo, Tokyo 113,
        Japan}
\\~\\
${}^{1}$\ Institut f\"ur Theoretische Physik,
Universit\"at zu K\"oln,
D--50937 K\"oln, Germany,
{\tt schreck@thp.uni-koeln.de} and {\tt as@thp.uni-koeln.de}
\\
${}^{2}$\ Zentrum f\"ur Paralleles Rechnen ZPR,
Universit\"at zu K\"oln,
D--50923 K\"oln, Germany,
{\tt kai@zpr.uni-koeln.de}
\\
${}^{3}$\ CISC JAERI,
Tokai, Ibaraki 319-11, Japan,
{\tt ito@catalyst.tokai.jaeri.go.jp}
}

\wp

\maketitle

\begin{abstract}
We investigate a probabilistic cellular automaton model which has been
introduced recently. This model describes single-lane traffic flow
on a ring and generalizes the asymmetric exclusion process models.
We study the equilibrium properties and calculate the so-called fundamental
diagrams (flow vs.\ density) for parallel dynamics. This is done numerically
by computer simulations of the model and by means of an improved
mean-field approximation which takes into account short-range correlations.
For cars with maximum velocity 1 the simplest non-trivial approximation gives
the exact result. For higher velocities the analytical results, obtained
by iterated application of the approximation scheme, are in
excellent agreement with the numerical simulations.
\end{abstract}



\section{Introduction}

Recently, there has been considerable interest in the investigation of
traffic flow using methods of statistical physics. Independently in
\cite{NS92} and \cite{BML} cellular automaton models \cite{Wo} for the
description of traffic flow have been proposed. Due to their computational
simplicity, lattice gas automata \cite{Wo} have already successfully
been applied to other problems, e.g. the simulation of fluids
\cite{FHP86} (for further applications, see the book of Wolfram \cite{Wo}).

A similar class of discrete models --- which may be interpretated as
traffic models or as models for surface roughening --- have also been
used for the description of the so-called asymmetric exclusion process
(driven diffusion) [5--13].  Here several exact solutions have been
obtained for processes where the particles can move at most one
lattice spacing per update step.  A natural generalization would allow
particles to move over larger distances. This is more realistic with
regard to modelling traffic since one usually has a whole spectrum of
allowed car velocities. These generalized exclusion models are thus
more appropriate for comparison with 'experiments' (i.e.~measurements
on freeway traffic \cite{HBG86}).

In addition there have been a number of publications dealing with
traffic flow in the framework of statistical mechanics, especially
using cellular automata \cite[15--17]{NS92,BML}. In most of
these works two-dimensional traffic has been studied which corresponds
to modelling the complex situation of traffic in the street-network of
a city.  Again the cars were allowed to move at most one lattice site
per time step. It could be shown numerically \cite{NS92} that relaxing
this constraint, i.e.\ the cars can have integer velocities up to an
upper speed limit larger than one, behaves qualitatively as well as
quantitatively in good agreement with real traffic (with an
appropriate upper speed limit). It will be shown in this paper that,
from an analytical point of view, the situation changes drastically
when turning from maximum velocity one to higher speed limits. In this
case real long-range correlations occur even in the stationary state
which is not true for models with maximum velocity one.

The exclusion models mentioned above may be classified according to the
boundary conditions and dynamics used. There are two relevant types of
boundary conditions: periodic and open. Using periodic boundary
conditions one considers a ring on which the cars/particles can move
('Indianapolis situation'). Open boundaries correspond to the so-called
'bottleneck situation' where one imposes certain input and output flows
at the chain ends.

Basically one has to distinguish four types of dynamics.  The dynamical
variables may be updated one after  the other in a certain order (sequential
update), one after another in random order (random-sequential), in
parallel for all sites of a given sublattice (sublattice update) or in
parallel for all sites (parallel update). In the case of asymmetric
update rules and sequential dynamics one has to distinguish at least two
cases: update in or opposite to the direction of the traffic motion.
Note that for open boundary conditions sequential and parallel
dynamics are identical if the update direction is the same as the direction
of traffic flow. For periodic boundary conditions, however,
they are not identical. Here parallel dynamics correspond to sequential
dynamics with a special site which closes the ring. This site memorizes
a car even if it moved away one time-step before. This creates
an obstacle in the ring giving rise to a lower flow through this
site compared with sequential dynamics in the direction of the motion.

In the present paper we use periodic boundary conditions and parallel
update. In this case not much is known exactly, since all of the
previous works~[5--13]
used either random-sequential or sublattice update. The advantage of
parallel update with respect to sublattice or sequential update is
that all sites are equivalent which should be the case in a realistic
model with translational invariance.  On the other hand parallel
update enhances the possible system sizes in numerical simulations,
especially because parallel or vector computers can be used easily.

The paper is organized as follows: In Sect.~2 we introduce the model.
In Sect.~3 the results of computer simulations are presented. We
describe the different simulation techniques which have been used and
compare their performance. In Sect.~4 we present a mean-field analysis
of the model for arbitrary velocities $\vm$. In Sect.~5 we introduce
the so-called $n-$cluster approximation~\cite{SS93}. This improved mean-field
method takes into account short-range correlations. In the case $\vm
=1$ we show that the $2-$cluster approximation already gives the exact
result. For $\vm =2$ we compare the results of the $n-$cluster
approximation ($n=1,\ldots 6$) with the computer simulations of
Sect.~3 and find an excellent agreement.  In App.~A the exact solution
for the case $\vm=1$ is derived whereas in App.~B the equivalence of
this case to an Ising model with repulsive interactions is shown.

\section{The model}

In the following we study single-lane traffic on a ring of length $L$ with
periodic boundary conditions. The model which has been introduced in
\cite{NS92} is defined as follows:\\
Each of the $L$ sites can either be empty or occupied by one vehicle with
velocity $v=0,1,\ldots ,\vm$. At each discrete time-step $t\to t+1$ an
arbitary arrangement of $N$ cars is updated according to the following
rules:
\bi

\item[1)]
{\bf Acceleration:} If the velocity~$v$ of a vehicle is lower than
$\vm$ the speed is advanced by one [$v=v+1$].

\item[2)] {\bf Slowing down (due to other cars):}
If the distance $d$ to the next car ahead is not larger than $v$ ($d \le v$)
the speed is reduced to $d-1$ [$v = d-1$].

\item[3)] {\bf Randomization:}
With probability $p$, the velocity of a vehicle
(if greater than zero) is decreased by one [$v=v-1$].

\item[4)] {\bf Car motion:}
Each vehicle is advanced $v$ sites.

\ei

These rules are applied to all cars in parallel (parallel update).
The rules ensure that the total number $N$ of cars is conserved under
the dynamics (which is not true in the bottleneck-situation). Note
that even for parallel update the randomization yields
non-deterministic behaviour. For random-sequential update the
probability $p>0$ is not essential because it
only rescales the time axis \cite{NS92}.

In the simplest case $\vm=1$ the cars are allowed to move only one step
during an update. For this situation several results are known
[5--13]. Especially it can be shown that for random-sequential
update the mean field Ansatz yields the exact equilibrium state
\cite{NS92,GS92} which is equivalent to the fact that for a fixed number of
cars every arrangement of cars occurs with the same probability. Therefore it
is quite natural to take the mean-field approach also as a starting point for
the investigation of higher velocities $\vm>1$ and parallel update.

Our main interest will be the calculation of the so-called fundamental diagram
(flow $q$ vs.\ density $\rho =N/L$). As described in \cite{NS92} these results
can be compared directly with measurements of real traffic \cite{HBG86}. One
expects a transition from laminar flow to start-stop waves with increasing car
density. For $\vm=1$ it is easy to see that the fundamental diagram is
symmetric with respect to $\rho=1/2$ due to particle-hole symmetry. This is not
true for realistic traffic where on finds a distinct asymmetry. i.e.\ the
maximum of the flow is shifted to lower values of $\rho$ ($\sim 0.2$).

The rules given above take into account the basic features of real
traffic. Firstly, they allow for a spectrum of velocities which seems to
be necessary in order to break the particle hole symmetry of the model
with only two velocities $\vm=1$. The maximum velocity still enters
the system as a free parameter but it can be argued \cite{NS92} that
most realistic are values around $\vm=5$. Secondly, the acceleration of the
cars is very smooth compared to the deceleration which can occur in
only one time step. For simplicity we assume the same maximum velocity
for all cars but this condition is not essential. Finally the
randomization step is necessary to avoid purely deterministic dynamics
and to take into account natural fluctuations of driving.

The order of the update rules given above is crucial.  Changing it
would also change the performance of the model. The randomization step
could e.g.\ be placed between acceleration and deceleration
(1--3--2--4) but then its influence is lowered drastically since every
decelerating car does not feel any randomness. On the other hand one
could also change the starting step. This will not influence the
properties of the model but can simplify the calculations.  If one
begins with step 2 (as we will do for the $n$-cluster approximation)
and proceeds 3--4--1 then one has the advantage that no cars with
velocity zero occur since all cars were just accelerated by one unit.
Therefore the number of possible states of a site is reduced by one.

{ 

\def\medt#1{\overline{#1}}

\def\epsfbox#1{\relax}
\def\epsfxsize#1{\relax}

\def\mmm#1{{\bf[[#1]]}}
\def\kai#1{\marginpar{$\longleftarrow$ KN?}\mmm{#1}}
\def\ms#1{\marginpar{$\longleftarrow$ MS?}\mmm{#1}}

\section{The simulations}

An integral part of our research were computer simulations.
Simulations allow to obtain quantitative insight into a model in
relatively short time.  In this way, they complement the analytical
work: Simulations are first used to test a variety of models until one
has some overview and the most useful is found, and high quality
simulation data are used to confirm analytical results (see
Figs.~\ref{V1} and~\ref{V2}).  Further on, simulations are used to go
beyond the analytically treated cases: either for
variations/extensions of the model, or for more complicated quantities
and issues such as the life-time distribution of the simulated traffic
jams~\cite{Nag94} or the possibility of self-organized
criticality~\cite{NaP94}.
Last, but definitely no least, this work are the first steps towards an
ultrafast microscopic simulation model for large traffic networks.  We
already have results for
2-lane-traffic~\cite{rickert.thesis,staatsexamen,2.lane.in.prep} and
for real world network implementations including ramps, intersections,
and junctions~\cite{rickert.thesis,alife,kai.thesis}.

\subsection{Simulation technicalities}

Initially, we used a simple code on a workstation (and sometimes its
replication on several nodes of a parallel computer) for getting an
overview over the model's properties and for estimating its relevance
for real world traffic~\cite{NS92}.  Later, we implemented
vectorizing and/or parallelized versions in order to obtain faster
simulation speeds.

Besides the advantage of getting high quality data in relatively short
time, we have taken advantage of the simplicity of the model to
implement it with different algorithms on many different
supercomputers.  This gives us intuition on how to implement more
complex models~\cite{TRANSIMS}, and to make predictions on how fast
our simulations will be in comparison to other microscopic traffic
simulation models~\cite{NaS94}.  In the cases of 2-lane-traffic and
of the network implementation, our predictions were quite accurate.

For the practical coding, we considered three different approaches:
site oriented, particle oriented, and an intermediate scheme.  {\it
Site oriented\/} directly implements the CA: A street is represented
by a chain~$\vec v$ of integers with values between $-1$ and
$v_{max}$, where $-1$ means that there is no particle at this site,
whereas the other values denote a particle and its velocity.  In
contrast, {\it particle oriented\/} means that two lists
$(x_i)_{i=1,\ldots,N}$ and $(v_i)_{i=1,\ldots,N}$ contain position
$x_i$ and velocity $v_i$ of each particle~$i$ ($i = 1, \ldots, N$).
This is similar to a molecular dynamics algorithm, except that
particles are constrained to integer positions and velocities.

Obviously, the particle-oriented approach will always be faster than
the site-oriented one for sufficiently low vehicle densities.  The
particle-oriented approach is more flexible, and since in single-lane
simulations passing of vehicles is not possible, the particle lists
are always ordered, making efficient codes for all kinds of computers
easy to write.  In addition, an extension to continuous position and
velocity is straightforward~\cite{NaH93}.

On the other hand, for the site-oriented (cellular automaton) approach
single-bit coding~\cite{Sta91} is possible. This means
that the model is formulated in logical variables, which may be stored
bitwise into computer words.  Logical operations on computer words
treat all bits of the word simultaneously, giving a theoretical
speedup of $b$, where $b$ is the number of bits per word (usually~32
or~64).  However, the {\it practical\/} gain for traffic simulations
on a workstation is much lower because the bit-oriented approach
cannot take advantage of the fact that only a fraction of all sites is
occupied by a particle.  Nevertheless, we found that, on a
workstation, the single-bit algorithm is faster for densities above
$0.05$ (for $v_{max}=5$).  In addition, the single-bit code runs very
efficiently on a Thinking Machines CM-5 using dataparallel CM-Fortran
and on a NEC-SX/3 traditional vector computer.  The simulation data
for the fundamental diagrams have been obtained this way.

Once passing of vehicles is allowed (multi-lane traffic), for the
particle oriented approach efficient memory allocation for parallel
and/or vector processors becomes more difficult, and single-bit coding
for the site oriented approach becomes tiresome.  These observations
led to a third, {\it intermediate\/} approach.  As in the site
oriented approach, each site is in one of $(v_{max}+2)$ states, but
for the update only the relevant sites are considered.  It turns out
(see below) that on parallel but not vectorizing computers this
algorithm is about as fast as the single-bit version.

In Table~\ref{cpu}, we give an overview of the computational speeds on
selected computers (see \cite{NaS94} for more details on most of these
results).  All values are valid for a vehicle density of $\rho = 0.1$
and system size of $L = 1\,333\,333$ sites, corresponding to 10\,000
single lane kilometers.  MUPS corresponds to Mega-Updates Per Second,
i.e., the number of sites updated per second divided by $10^6$.  These
values are useful in order to compare with other implementations of
similar cellular automata or particle hopping systems.  The other
number is the extrapolated real time system size, i.e.\ the
extrapolated system size (assuming linear speed-up) where the
computation would be just as fast as reality.

The most noteworthy features of the table are emphasized in
boldface:\begin{itemize}

\item On vectorizing computers such as the SX-3 and the CM-5, the
single-bit algorithm has a notable advantage over the other
algorithms.  On all other machines, the intermediate algorithm is
never more than a factor 2.5 slower than the single-bit algorithm.

\item Already on a relatively modest machine such as an Intel Paragon
with 64~nodes, our real time limit is~280\,000~single lane kilometers.
Since, e.g., the whole freeway network of Germany is about
60\,000~single lane kilometers long (12\,000~km $\times$ 2~directions
$\times$ 2.5~lanes), the real use of this computational speed will be
(i)~real time applications, where the traffic forecast has to be
computed before the situation arrives; and (ii)~Monte Carlo
simulations, where many runs are necessary.

\item
The by far fastest ``realistic'' traffic microsimulation world-wide to
  date is the PARAMICS microsimulation project~\cite{PARAMICS}.  Their
  real time limit is $\approx$ 20\,000~km on 16k CM-200, a machine
  which is in terms of peak performance slightly faster than a
  128-node-Paragon.  In other words, it seems that our not completely
  realistic car following logic buys us about an order of magnitude in
  performance.

\end{itemize}

\subsection{Simulation results}

Figs.~\ref{V1.pixel} and~\ref{parallel.pixel} show typical time
evolutions of different versions of the model.  Fig.~\ref{V1.pixel}
shows parallel update with $v_{max}=1$ on the left, and random
sequential update with $v_{max}=5$ on the right, both at density
$c=0.5$.  Random sequential update for $v_{max}=1$ looks the same as
the left picture.  Obviously, neither using parallel update for
$v_{max}=1$ nor taking a $v_{max}>1$ for random sequential update does
change the phenomenlogical behavior from the stochastic asymmetric
exclusion process.  For $v_{max}=1$, $\rho=0.5$ corresponds to maximum
flow, and in consequence, wave structures do not move in
space~\cite{Krug}.  For $v_{max}>1$, the point of maximum throughput
is shifted to lower densities, and in consequence, in the left plot
the waves are moving backwards.

The situation is different when one combines parallel update and
$v_{max}>1$ (Fig.~\ref{parallel.pixel}).  Here, in the regime of
maximum throughput (left plot) waves are only sparse, and they clearly
move backwards.  And even at higher densities (right plot), waves are
much more distinct than in the random sequential case.

In short, one can divide the models between random sequential update
with arbitrary maximum velocity on one hand, and parallel update with
maximum velocity $v_{max} \ge 2$ on the other hand.  Parallel update
with $v_{max}=1$ phenomenologically is an intermediate case.
Interestingly, it will turn out that this structure is reflected in
the analytical calculations below: For random sequential update, the
mean field solution is already exact.  For parallel update and
$v_{max}=1$, the situation is only slightly different because already
the first step beyond mean field is exact.  The situation is
completely different for higher velocities in connection with parallel
update, where the analytical approximations only converge slowly
towards the simulation result.

In addition, Fig.~\ref{parallel.pixel} allows an interesting
comparison with fluid-dynamical models.  Starting from ordered initial
conditions, one clearly sees how instabilities develop and produce the
start-stop-waves, very similar to results in~\cite{Kerner.Konh}.
Working out these connections is the topic of current
research~\cite{kai.thesis}.

Fig.~\ref{fdiag}~(a) shows current vs.~density curves for maximum
velocities $v_{max}$ between 1 and 5, plus for a different fluctuation
parameter $p=0.25$ at $v_{max}=5$.  One clearly sees that the maximum
throughput increases with increasing $v_{max}$, whereas the density of
maximum throughput decreases.  In reality, the density of maximum
throughput lies between $\rho = 0.15$ and $\rho = 0.2$; it is given by
the maximum speed of trucks which dominate the speed distribution for
traffic near capacity~\cite{rickert.thesis}.  For that reason, having
a higher speed limit for passenger cars does not help throughput; in
many cases, it actually makes things worse~\cite{Zackor.Kuehne.Balz}.

However, reducing the fluctuation parameter $p$ increases throughput
enormously.  This is mostly due to the better acceleration behavior in
that case~\cite{trb}.

In general, one sees that by varying the parameters $v_{max}$ and $p$,
the fundamental diagram can easily be adapted to real traffic
situations, although some of the underlying vehicle dynamics remain
somewhat unrealistic: E.g.\ average acceleration from 0 to 100~km/h
takes place in 10~seconds.  That fact that the fundamental diagram is
nevertheless quite realistic is due to the fact that the first time
steps fo the acceleration matter most~\cite{Piper}.  And here,
4~seconds for an acceleration from~0 to 40~km/h are far more
realistic.

Another quantity of interest for traffic engineers are the velocity
fluctuations
\[
\sigma(v_{loc}) := \sqrt{  \medt{ ( v_{loc} - \medt{v_{loc}} )^2 } } \ ,
\]
where $v_{loc}$ is the ``local'' velocity of vehicles crossing a fixed
line.  (The average of the local velocity is different from the usual
ensemble average.  For example, cars with velocity zero never enter
the local average.)  According to measurements and fluiddynamical
arguments~\cite{Zackor.Kuehne.Balz}, these fluctuations are a good
indicator of traffic near capacity.  And indeed do we find for our
model (Fig.~\ref{fdiag}~(b)) that these fluctuations very abruptly
increase near capacity.

} 

\newcommand\vmax{v_{\rm max}}
\newcommand\dit[2]{d(#1,#2)}
\newcommand\cilt[3]{c_{#2}(#1,#3)}
\newcommand\cit[2]{c(#1,#2)}
\newcommand\dt[1]{d(#1)}
\newcommand\clt[2]{c_{#1}(#2)}
\newcommand\ct[1]{c(#1)}

\section{Mean-Field Theory}

The complete analytic solution of the traffic model is not possible
except for the case of maximum velocity $\vmax =1$ (see Sect.~5.1 and
Appendix A). Therefore some approximation is necessary when one tries
to study this model analytically for $\vmax > 1$.  The following
mean-field type theory will be the first step of such an analytical
approach.

The calculation starts from the stochastic description of the traffic
model. Instead of specifing the car position and their velocities, we
analyze the probability distribution of each site at each time step.
We denote the probability that there is no car at site $i$ ($i=1$,
$2$, $3$, $\cdots$, $L$) at timestep $t$ by $\dit{i}{t}$ and the
probability that there is a car with velocity $\alpha$ ($\alpha =0$,
$1$, $2$, $\cdots$, $\vmax$) at site $i$ and timestep $t$ by
$\cilt{i}{\alpha}{t}$.  The $c$ and $d$ distributions together take
into account all possible states of the different sites.  Therefore
one has the normalization condition for all sites and all timesteps
\begin{equation}
\dit{i}{t} +
\cilt{i}{0}{t} + \cilt{i}{1}{t} +
\cilt{i}{2}{t} + \cilt{i}{3}{t} + \cdots
\cilt{i}{\vmax}{t} =1\; .
\label{EQPROBCONSERVATION}
\end{equation}
Denoting with $\cit{i}{t}$ the total probability
for site $i$ to be occupied at timestep $t$, i.e.\
$\sum_{j=0}^{\vmax}\cilt{i}{j}{t}$, one simply has $
\dit{i}{t} + \cit{i}{t} =1$.

According to the update rules in four stages (see Sect.\ 2) the
time evolution of these probability distributions can be described by
the following four sets of equations:
\begin{itemize}
\item{Acceleration Stage}
\begin{eqnarray}
\cilt{i}{0}{t_1} &=& 0\nonumber\\
\cilt{i}{\alpha}{t_1} &=& \cilt{i}{\alpha -1}{t}\;,
                                                \quad\quad 0 < \alpha < \vmax\\
\cilt{i}{\vmax}{t_1} &=& \cilt{i}{\vmax}{t}+\cilt{i}{\vmax -1}{t}\nonumber
\end{eqnarray}
\item{Deceleration Stage}
\begin{eqnarray}
\cilt{i}{0}{t_2} &=& \cilt{i}{0}{t_1}
         +\cit{i+1}{t_1}\sum_{\beta =1}^{\vmax}\cilt{i}{\beta }{t_1}\nonumber\\
\cilt{i}{\alpha}{t_2} &=& \cit{i+\alpha+1}{t_1}
           \prod_{j=1}^{\alpha}\dit{i+j}{t_1}
            \sum_{\beta=\alpha+1}^{\vmax}\cilt{i}{\beta}{t_1}\\
      &+&\cilt{i}{\alpha}{t_1}
      \prod_{j=1}^{\alpha}\dit{i+j}{t_1},\quad 0 < \alpha < \vmax\nonumber\\
      \cilt{i}{\vmax}{t_2} &=& \prod_{j=1}^{\vmax}
                         \dit{i+j}{t_1}\cilt{i}{\vmax}{t_1}\nonumber
\end{eqnarray}
\item{Randomization Stage}
\begin{eqnarray}
\cilt{i}{0}{t_3} &=& \cilt{i}{0}{t_2} + p\cilt{i}{1}{t_2}\nonumber\\
\cilt{i}{\alpha}{t_3} &=&
           q\cilt{i}{\alpha}{t_2} + p\cilt{i}{\alpha +1}{t_2},\quad
           0 < \alpha < \vmax\\
\cilt{i}{\vmax}{t_3} &=& q \cilt{i}{\vmax}{t_2}\nonumber
\end{eqnarray}
\item{Motion Stage}
\begin{eqnarray}
\cilt{i}{\alpha}{t+1} &=& \cilt{i-\alpha}{\alpha}{t_3},
            \quad 0\le\alpha \le\vmax
\end{eqnarray}
\end{itemize}
The time $t$ is assumed to take on only integer values. $t_1$, $t_2$
and $t_3$ denote the intermediate time steps between $t$ and $t+1$
(sometimes defined as $t+1/4$, $t+1/2$ and $t+3/4$,
respectively). These time-evolution equations conserve for
periodic boundary conditions the total number of
cars at each stage . This formulation of the dynamics neglects spatial
correlations completely since one assumes that all expectation
values factorize into local terms.

The variables $c$ and $d$ are real valued between $0$
and $1$. The stochastic description originates from the stochastic
nature of the third randomization step. The other three steps are
purely deterministic.

These time evolution equations are non-linear and further analysis of
the full system has not been successful up to now.  However, in the
stationary state, i.e.\ in the limit $t\to\infty$, the $c$ and $d$
distributions become homogeneous in space (for periodic boundary
conditions) and therefore, apart from the time dependence, also the
site dependence can be omitted. Using this and combining the four
update steps one gets the following set of $\vmax+1$ equations:
\begin{eqnarray}
c_0 &=& (c+pd) c_0
    + (1+pd)c\sum_{\beta = 1}^{\vmax} c_{\beta}\nonumber\\
c_{\alpha} &=& d^\alpha[ qc_{\alpha -1} +
(qc+pd)c_{\alpha} + (q+pd)c\sum_{\beta = \alpha +1}^{\vmax}
c_{\beta}], \\
& & \quad\quad\quad\quad\quad 0 < \alpha < \vmax-1  \nonumber\\
c_{\vmax -1} &=& d^{\vmax -1}[qc_{\vmax -2} +
(qc+pd)(c_{\vmax -1}+ c_{\vmax})] \nonumber\\
c_{\vmax} &=& qd^{\vmax}[c_{\vmax-1} + c_{\vmax} ]\nonumber
\end{eqnarray}
These equations essentially describe the motion of a single car with
statistical (densitity dependent) 'obstacles'.  A remarkable feature
of the equations is that they are linear when one specifies the
density $c=1-d$ of cars.  Therefore (4.6) can be recast in matrix form
as ${\bf A}\vec c = \vec c$. The matrix ${\bf A}$ can be read off from
(4.6), $\vec c$ is the vector with elements $c_\alpha$,
$\alpha=0,...,\vm$. For small $\vm$ we can invert ${\bf A}$ to find the
densities $c_\alpha$ explicitly. For large values of $\vm$ it is more
convenient to write down a recursion relation in order to obtain the
steady state solution.

 From the first equation in (4.6) one can determine
$c_0$ directly without knowledge of the other $c_{\alpha}'s$ to give
\begin{equation}
c_0 = c^2 {1+pd\over 1-pd^2}.
\end{equation}
Using this result and the second equation of (4.6) one can also write
down the expression for $c_1$
\begin{equation}
c_1 = qc^2 d {1+d+pd^2\over (1-pd^3)(1-pd^2)}.
\end{equation}
For $\alpha$ larger than one a recursion relation can be derived
calculating $c_\alpha - d c_{\alpha-1}$ using
again the second equations of (4.6)
\begin{equation}
c_{\alpha} = {1+(q-p)d^{\alpha}\over 1-pd^{\alpha +2}}dc_{\alpha -1}
- {qd^{\alpha} \over 1-pd^{\alpha +2} } c_{\alpha -2}
\end{equation}
for $\alpha = 2$, $3$, $\cdots$, $\vmax -2$. Therefore, starting
with the expressions (4.7), (4.8) for $c_0$ and $c_1$, one can
estimate $c_2$, $c_3$, $\cdots$, $c_{\vmax -2}$ recursively.
These results do {\it not} depend on the actual value of $v_{max}$
and thus are valid generally (provided $v_{max}\ge 2$).

Finally, from the last two equations of (4.6) one gets
\begin{eqnarray}
c_{\vmax -1} &=&{1-qd^{\vmax} \over 1-d^{\vmax-1}(q+pd)}
  qd^{\vmax-1}c_{\vmax -2} \\
c_{\vmax} &=&{qd^{\vmax} \over 1-qd^{\vmax}}c_{\vmax -1}.
\end{eqnarray}
The $\vmax$ dependence only occurs in these two quantities.  In
Fig.~\ref{rho_v} some of the densities are shown for large $\vmax$.
The densities of the fast cars go to zero rapidly, since one expects
an exponentially fast decay from the recursion relations,
(4.9)--(4.11). The contributions from cars with high velocities
therefore are neglegible.  We will mainly be interested in the flow
$f(c,p)$, defined by
\begin{equation}
f(c,p) = \sum_{\alpha =1}^{\vmax} \alpha c_{\alpha}.
\end{equation}
In the limit of $\vmax \rightarrow \infty$ it is possible
to carry on the analysis using the iteration equations (4.7)--(4.9)
and the generating function
\begin{equation}
g(z)=\sum_{\alpha =0}^{\infty} z^\alpha c_{\alpha}.
\end{equation}
As usual one simply has $g'(1)=f(c,p)$. The equations (4.7)--(4.9)
now can be combined to give one single equation for $g(z)$
\begin{equation}
g(z)[1-dz]-g(dz)d^2(1-z)[p+qz]=c^2+pc^2d(1-z).
\end{equation}
Since $\vmax$ is infinite one does not have to worry about the upper
boundary equations (4.10) and (4.11) for $c_{\vmax-1}$ and $c_{\vmax}$
whose contribution is neglegible.  One should notice that the
generating function $g$ occurs with two different arguments ($z$ and
$dz$) which makes it impossible to solve this (linear) equation for
$g(z)$ explicitly.

After differentiation an expression for $g'(1)$ is obtained
\begin{equation}
g'(1)=d(1-pc)-g(d)\frac{d^2}{c}
\end{equation}
The problem therefore is reduced to find an expression for $g(d)$.
This can be found by successive application of equation (4.14) with
$z=d^n$, $n=1,2,3,\ldots$. The final result is an asymptotic
expression for $g'(1)=f(c,p)$
\begin{equation}
f(c,p)=qcd\left[ 1+\sum_{n=1}^{\infty}d^{2n}\prod_{l=0}^{n-1}(p+qd^l)\right]
\end{equation}
In Fig.~\ref{flow.inf} the flow $f$ is shown as a function of the
concentration $c$ of cars for various deceleration probabilities $p$.
The slope at the origin ($c\sim 0$) of the fundamental diagram is
infinite whereas the slope for $c\sim 1$ is $-q$. This mean-field
result yields, compared with the simulation data shown above, much too
small values of the flow. This can easily be understood since the
reduction to a single car problem ignores all spatial correlations of
the cars.  Cars with high velocities tend to be equidistant and can
therefore maintain a high velocity with a larger probability than
in the mean-field system where it is much more difficult to accelerate
and stay at high velocities over a certain period.

Furthermore, even for $\vm=1$, the mean-field solution
does not coincide with the exact result shown below. Note that for
random sequential update the mean-field solution is exact for
$\vm=1$.

\setcounter{equation}{0}

\section{Improved Mean-Field Theories}

In order to improve the simple mean-field theory of the preceeding section
we have to take into account correlations between neighbouring sites
\cite{SS93}.

We divide the lattice into segments or clusters of length $n$
($n=1,2,\ldots$) such that two neighbouring clusters have $n-1$ sites
in common.  The probability to find in equilibrium a cluster in state
$(\sigma_1,\ldots,\sigma_n)$ will be denoted by
$P_n(\sigma_1,\ldots,\sigma_n)$. Due to the translational invariance of
equilibrium state of the system with periodic boundary conditions one
has not to specify the actual location of the $n$-spin cluster.  In
order to simplify the calculations we apply, as mentioned above, the
four update-rules in the order 2--3--4--1 instead of 1--2--3--4
\cite{SS93}.  This has the advantage that after one update cycle one
ends up with step 1 and therefore no car has velocity 0. It follows
that every site $j$ is in one of the $\vm$ states
$\sigma_j=0,1,\ldots,\vm$ where now 0 denotes an empty site. This
change in the ordering finally has to be taken into account in the
calculation of the flow.

The equilibrium probabilities for an $n$-site cluster are then
determined by
\begin{eqnarray}
P_n(\sigma_1,\ldots,\sigma_n) & = & \sum_{\{\tau_j\}}
W(\sigma_1,\ldots,\sigma_n\vert\tau_{-\vm +1},\ldots,\tau_{n+\vm})
\times\nonumber\\
& & \makebox[3.5cm]{}\times
P_{2\vm+n}(\tau_{-\vm+1},\ldots,\tau_{n+\vm})\, ,
\label{eqeq}
\end{eqnarray}
where the probability $P_{2\vm+n}(\tau_{-\vm+1},\ldots,\tau_{n+\vm})$
for clusters of length $2\vm+n$ has to be expressed through the
$n$-cluster probabilities $P_n(\tau_1,...,\tau_n)$. This enlargement of
the cluster length occurs since all cars which can drive into or out
of the cluster $(\sigma_1,\ldots,\sigma_n)$ within the next timestep
contribute to the transition rates $W(\{\sigma_j\}\vert\{\tau_j \})$.
Thus we have to take into account not only the given cluster but also
the $\vm$ sites to its left (with state variables
$\tau_{-\vm+1},\ldots,\tau_0$) and the $\vm$ sites to its right (with
state variables $\tau_{n+1},\ldots,\tau_{n+\vm}$).
The decomposition of the $(2\vm+n)$-cluster can be carried out
by introducing the conditional probabilities
\begin{equation}
P(\tau_i\vert\underline{\tau_{i+1},\ldots,\tau_{i+n-1}})=
{P_n(\tau_i,\tau_{i+1},\ldots,\tau_{i+n-1})\over \sum_{\tau}
P_n(\tau,\tau_{i+1},\ldots,\tau_{i+n-1})}
\label{cprobr}
\end{equation}
at the left side and
\begin{equation}
P(\underline{\tau_i,\ldots,\tau_{i+n-2}}\vert\tau_{i+n-1})=
{P_n(\tau_i,\tau_{i+1},\ldots,\tau_{i+n-1})\over\sum_{\tau}
P_n(\tau_i,\ldots,\tau_{i+n-2},\tau)}
\label{cprobl}
\end{equation}
at the right side. With this definition we rewrite $P_{2\vm+n}$ (in the
$n$-cluster approximation) in the following form:
\begin{eqnarray}
P_{2\vm+n}(\tau_{-\vm+1},\ldots,\tau_{n+\vm})
 & = &\prod_{i=-\vm +1}^{0}
P(\tau_i\vert\underline{\tau_{i+1},...,\tau_{i+n-1}}) \times\label{trans}
\\
 & & \mbox{}\times P_n(\tau_1,\ldots,\tau_n)\prod_{i=1}^{\vm}
P(\underline{\tau_{i+1},...,\tau_{i+n-1}}\vert \tau_{i+n})\, .\nonumber
\end{eqnarray}
The transition probability $W(\sigma_1,\ldots,\sigma_n\vert\tau_{-\vm
+1},\ldots,\tau_{n+\vm})$ depends on the probability $p$ and vanishes
if the configuration $(\tau_{-\vm+1},\ldots,\tau_{n+\vm})$ cannot
evolve in {\em one} timestep into $(\sigma_1,\ldots,\sigma_n)$
according to the rules $1)-4)$ of Sect.~2. If $W$ is non-zero it is of
the form $p^{n_1}q^{n_2}$ with integers $n_1,n_2$ describing how many
cars have to be decelerated ($n_1$) through the randomization step
when total number of cars which can drive is $n_1+n_2$. With the
approximation (\ref{trans}) it is possible to write down a closed
system of equations for the $n$-cluster probabilities
$P_n(\sigma_1,\ldots,\sigma_n)$. The number of the equations is given
by $(\vm+1)^n$, the total number of possible configurations of $n$
site variable with $\vm+1$ possible states\footnote[1]{In practice
some of these equations turn out to be trivial so that the relevant
number is less than $(\vm+1)^n$.} (without change of the order of the
update steps one would have $(\vm+2)^n$ equations). Due to the
exponential growth with respect to $n$ one is, especially for larger
$\vm$, restricted to only small cluster lengths $n$ (For the realistic
value of $\vm=5$ one has for the 2-cluster approximation already 36
equations!)

The above approximation converges for $n\to\infty$ to the exact
solution for an infinite system (i.e.\ in the thermodynamic limit
$L\to\infty$. This approximation scheme is well known in the
literature.  It is analogous to the $(n,n-1)-$cluster approximation of
\cite{BEN92}, the $n-1$ step Markovian approximation of \cite{CRIS} or
the local structure theory of \cite{GUTO}. It goes back to the
probability path method introduced by Kikuchi \cite{KIKU}.

With the knowledge of the $n$-cluster probabilities
$P_n(\sigma_1,\ldots,\sigma_n)$ it is then easy to calculate the
fundamental diagram, i.e.\ the flow $f$ as a function of the density
$c$ of cars. Since we have changed the order of the update steps one
has to take this into account by performing the steps 2--3--4 at the
end since the last step must be 4 ($\equiv$ car motion).
After that one simply can calculate the density $c_{\alpha}$ of cars
which will drive $\alpha$ sites in the next timestep
\begin{equation}
c_{\alpha}=\sum_{\sigma_2,\ldots,\sigma_n}P_n(\alpha,\sigma_2,\ldots,\sigma_n)
\end{equation}
and then one proceeds as in the mean-field approximation eqn.\
(4.12) and calculates $f=\sum\alpha c_{\alpha}$.

\subsection{{\bf $\vm = 1$}}

In the case $\vm=1$ the site variables take on only the values
$\sigma=0,1$ where $\sigma=0$ means no car and $\sigma=1$ a car with
velocity one. In the $2$-cluster approximation ((\ref{eqeq}) with
$n=2$) one has, according to the above arguments, a system of 4
equations. This can be be reduced to only one equation for
$P(1,0)$ very easily with the help of the relations
\begin{eqnarray}
P(1,0)&=&P(0,1)\nonumber\\
P(0,0)&=&1-c-P(1,0),\\
P(1,1)&=&c-P(1,0).\nonumber
\end{eqnarray}
The first equation is due to the 'particle-hole'
symmetry $P(1,0)=P(0,1)$ (in a closed ring one must have the
same number of (0,1) and (1,0) pairs, therefore occuring with
the same probability). The other two equations describe the conservation
of cars in the system. The remaining equation for $P(1,0)$ reads
\begin{equation}
qP^2(1,0)-P(1,0)+c(1-c)=0.
\end{equation}
In the thermodynamic limit we therefore find \cite{SS93}
\begin{equation}
P(0,1)=P(1,0)={1-\sqrt{1-4qc(1-c)}\over 2q}\ .
\label{prob}
\end{equation}
Going to the 3- and higher-cluster approximations one finds that the
solution remains the same indicating that this is the exact result. In
App.~A we indeed prove that the solution (\ref{prob}) is exact in the
thermodynamic limit.  For finite systems the prove is also valid but
one has to take into account an additional correction (normalization)
due to the constraint of a fixed number $N$ of cars.

The correct result, valid for any system size, is is given by
\begin{equation}
{\cal P}(N,L)={1\over {\cal{N}}}\sum_{\{\sigma\}}\,\hskip-4pt ' \prod_{j=1}^L
P(\sigma_j,\sigma_{j+1})
\label{ground}
\end{equation}
where ${\cal{N}}$ denotes the normalization constant and the sum $\sum
'$ runs over all configurations with a fixed number $N$ of cars (i.e.~
$\sum_{j=1}^L\sigma_j=N$).

In contrast to random-sequential dynamics parallel dynamics leads to
an effective attraction between 'particles' and 'holes' (i.e.
$P(0)P(1)=c(1-c) \le P(0,1)$) and thus to a higher flow.  In App.\ B
the mapping of this model to an equivalent Ising-model with
antiferromagnetic next-nearest-neighbour interactions and an
nonvanishing external field is shown. Due to the antiferromagnetic
interactions the system shows an effective attraction of cars over two
lattice sites, thus taking into account the well known effect of car
bunching or platooning~\cite{May,Ben-Naim,Nagatani-bunching}.

In order to calculate the flow one first has to perform
the steps 2--3--4 yielding the probabilities $\tilde{P}(\tau)$
to find a site in the state $\tau = -1,0,1$ (where now an empty
site is denoted by $\tau=-1$ ) after the {\em fourth} step of the updating
procedure:
\begin{equation}
\tilde{P}(-1)=1-c, \qquad \tilde{P}(0)= qP(1,1),\qquad \tilde{P}(1)= qP(0,1)
\label{fourstep}
\end{equation}
Therefore the fow is finally given by
\begin{equation}
f(p,c)=qP(1,0)={1-\sqrt{1-4qc(1-c)}\over 2}.
\label{flux}
\end{equation}
In Fig.\ 6 the exact result for the flow is shown (from the 2-cluster
approximation) in comparison to numerical simulations and the mean-field
approximation ($p=1/2$). One can see that the numerical and analytical
data are in excellent agreement.
In the deterministic case
$p=0$ the flow is a linear function of $c$: $f=(1-\vert1-2c\vert)/2$
\cite{SS93}.

The mean velocity per car $\bar{v}$ is then
\begin{equation}
\bar{v}={1\over \rho}\sum_{\tau=0}^{\vm} \tau\tilde P(\tau) =\tilde P(1)
/\rho\ .
\label{vmean}
\end{equation}
In the deterministic case $p=1$ this simplifies to
\begin{equation}
\bar{v}=\cases{1 &for $0\leq c\leq 1/2$\cr
(1-c)/c &for $1/2\leq c\leq 1$\cr}
\label{vmeanp1}
\end{equation}
which is the result of \cite{BML}.

\subsection{{\bf $\vm = 2$}}

The case $\vm =2$ is qualitatively very different from the case $\vm
=1$.  The flow diagram is no longer symmetric around $c=1/2$.  The
$n-$cluster approximation seems not to become exact for any finite
$n$, i.e.~in this case long-ranged correlations are important. We have
calculated the fundamental diagram for $n=1,\ldots,5$. As shown in
Fig.\ 7 the approximation converges fast to the results obtained
by simulations. The difference between $n=4$ and $n=5$ is less then
1\%.

The observation that the approximation does not become exact for small
$n$ reflects the fact that the physics for $\vm\ge 2$ is distinctly
different from the case $\vm = 1$. As explained in Sec.\ 2 the regime
looks qualitatively different both from $\vm = 1$ (arbitrary update)
as well as from random sequential update (arbitrary $\vm$). Moreover,
in the case $\vm\ge 2$ (parallel update), jams show characteristic
branching behavior which is not observed for $\vm = 1$ \cite{NaP94}.

\section{Conclusions}
We have introduced and investigated a statistical model capable to
describe accurately real traffic. Through the introduction of higher
velocities it was possible to produce the asymmetric fundamental
diagrams and the characteristic start-stop waves observed in real
traffic. Through simulation, different regimes depending on the type
of update and the maximum velocity have been identified.  Realistic
for traffic is a model with parallel stochastic update and a suitable
choice of the maximum velocity $\vm= 5$. Already for $\vm = 1$ the
stationary state is more complicated than for random sequential
update.  An effective 'antiferromagnetic' interaction between cars
favors equal spacing and in consequence higher throughput.

Taking into account two-point correlations the case $\vm=1$ can be
solved exactly. This is no longer true for higher maximum velocties
where correlations become long ranged. Nevertheless, it could be shown
that the $n$-cluster approximation converges fast to the simulation
data.


\acknowledgements

This work was performed within the SFB 341 K\"oln-Aachen-J\"ulich
supported by the DFG.  KN is supported by the ``Graduiertenkolleg
K\"oln/St.~Augustin''.  The Center for Parallel Computing ZPR K\"oln,
the German National Research Center KFA J\"ulich, the
``Rechenzentrum'' of the University of Cologne, and the ``Institut
f\"ur Angewandte Informatik'' of the University Wuppertal provided
computer time.

\appendix
\def\theequation{\thesection.\arabic{equation}}
\setcounter{equation}{0}
\section{Exact solution for $v_{max}=1$}
\newcommand{\sbf}{\mbox{\boldmath$\sigma$}}
\newcommand{\tbf}{\mbox{\boldmath$\tau$}}
In this Appendix we show that the stationary state of the
model with $\vm=1$ is in fact given by equation (5.6) with
an appropriate normalization constant {\em Z}.
The complete set of evolution equations for parallel update reads:
\formel
{P_{t+1}(\{\sbf\})=\sum_{\{\tbf\}} W(\{\sbf\}\,|\,\{\tbf\})
\cdot P_t(\{\tbf\}),}
where $P_t(\{\sbf\})$ denotes the probability for state
$\{\sbf\}=\{\sigma_1,\ldots,\sigma_L\}$ at time~$t$. The transition
probability $W(\{\sbf\}\,\vert\,\{\tbf\})$ to move in one timestep
from state $\{\tbf\}$ to state $\{\sbf\}$ factorizes into local terms:
\formel
{W(\{\sbf\}\,\vert\,\{\tbf\})=\prod_{i=1}^L
W(\sigma_i,\sigma_{i+1}\,\vert\,\tau_i,\tau_{i+1}).}
The only non-vanishing transition probabilities are given by:
\formarr{
W(0,1\,\vert\,1,0) & = & 1-p \nonumber\\
W(1,0\,\vert\,1,0) & = & p \nonumber\\
W(1,\sigma'|\sigma,1) & = & 1 \\
W(\sigma,0|0,0) & = & 1 \nonumber\\
W(0,\sigma|0,1) & = & 1 , \quad \sigma,\sigma'=0,1\nonumber}
whereas $W(\tau,\tau'\,\vert\,\sigma,\sigma') = 0$ in all other
cases.
(Note that in eqn.\ (5) of ref.\ \cite{NS92} due to a misprint a
factor $(1-\sigma_{i+1})/2$ is missing on the right-hand side.)
We now make the Ansatz that the probability in the stationary
state $P(\{\sbf\})$ factorizes into local two-site terms
$P_{\sigma_i\sigma_{i+1}}$:
\formel
{P(\{\sbf\}) = \prod_{i=1}^LP_{\sigma_i\sigma_{i+1}}}
with periodic boundary condition $\sigma_{L+1}=\sigma_1$.  We will
define $n_{\sigma,\sigma'}(\sbf)$ as the number of pairs of next neighbours
$(\sigma,\sigma')$ in a particular state ${\sbf}$. Due to the particle-hole
symmetry of the system one has $n_{01} = n_{10}$ and the following
simple relations for the system size $L$ and the (conserved) number of
particles $N$ hold:
\formarr{
L & = & 2\cdot n_{01}+n_{11}+n_{00}\\
N & = & n_{01}+n_{11}.}
In the stationary state equation (A.1) becomes time-independent and
the states $\{\tbf\}$ in the summation on the right-hand side can be
classified according to the number of particles $l$ which have to be
moved in order to obtain the new state $\sbf$:
\formarr
{\lefteqn{
(P_{01}P_{10})^{n_{01}}P_{11}^{n_{11}}P_{00}^{L-2n_{01}-n_{11}}=}\\
& &\sum_{l=0}^{n_{01}}\sum_{\Delta}g(n_{01},l,\Delta)(1-p)^lp^{n_{01}-l-\Delta}
(P_{01}P_{10})^{n_{01}-\Delta}P_{11}^{n_{11}+\Delta}
P_{00}^{L-2n_{01}-n_{11}+\Delta}.\nonumber}
The summation index $\Delta$ is defined as $\Delta =
n_{11}(\tbf)-n_{11}(\sbf)$. Therefore the range of the
$\Delta$-summation depends on the particular state $\sbf$.  The
function $g(n_{01},l,\Delta)$ counts the number of possible states
$\tbf$ leading to state $\sbf$ given fixed values of $n_{01}$, $l$ and
$\Delta$ and reflects a kind of degeneracy.  Summing $g$ over $\Delta$
yields simply
\formel
{\sum_{\Delta}g(n_{01},l,\Delta)={n_{01}\choose l}}
Dividing eqn.\ (A.7) by the left-hand side one gets
\formel
{1=\sum_{l=0}^{n_{01}}\sum_{\Delta}g(n_{01},l,\Delta)(1-p)^lp^{n_{01}-l}
\left(\frac{P_{00}P_{11}}{pP_{01}P_{10}}\right)^{\Delta}.}
Setting the expression in the parenthesis on the right-hand side to 1,
(A.9) reduces with the help of (A.8) to an identity. Therefore the
condition for $P(\sigma,\sigma')$ reads
\formel
{P_{00}P_{11}=pP_{01}P_{10}}
Note that $P(\sigma,\sigma')$ is not normalized for a {\it finite}
system and therefore a normalization constant $\cal N$ has to be
taken into account.

\setcounter{equation}{0}
\section{Mapping to an equivalent Ising-model}
Introducing Ising-variables $\sigma_i=\pm 1$ instead of the
lattice-gas variables $\tau_i=0,1$ ($\sigma_i = 2\tau_i-1$) one can
look at the steady state as the equilibrium distribution $P(\sbf)\sim
e^{-\beta H(\sbf)}$ of an Ising model with Hamiltonian
$H=-J\sum_i\sigma_i\sigma_{i+1} + h\sum_i\sigma_i$ In order to
determine the coupling constant $J$ and the external field $h$ one
interpretes the local probabilities $P_{\sigma,\sigma'}$ as transfer
matrices $P_{\sigma,\sigma'} = \alpha
e^{-J\sigma\sigma'-h(\sigma+\sigma')/2}$, ($\beta\equiv 1$). From
eqn.\ (A.10) it follows directly that $e^{4J}=p$ or $J={\rm
log}(p)/4<0$. Therefore the corresponding Ising model has
antiferromagnetic interactions.  According to eqns.\ (A.5) and
(A.6)
one has in addition
\formarr{ 1 & = & 2\cdot P_{01}+P_{11}+P_{00}\\
\rho & = & P_{01}+P_{11}.}
Dividing the two expressions on both sides of the two equations
gives one eqn.\ without the constant $\alpha$ to determine the
external field as
\begin{equation}
e^h={\sqrt{1-4(1-p)\rho(1-\rho)}-(1-2\rho)\over 2\sqrt{p}(1-\rho)}.
\end{equation}
Therefore the steady state corresponds to an
Ising-model with {\em antiferromagnetic} (repulsive) interactions.
Due to the conservation of the total number of particles (cars) one
has to impose the constraint of a fixed magnetization to the
Hamiltonian.  The normalization is then simply the partition function
calculated under this constraint.

\def\bfig{\begin{figure}[t]
}
\def\efig{\end{figure}}

{ 

\begin{table}
{ \baselineskip6pt \openup0pt \lineskiplimit0pt \tightenlines

\def\btable{
  \bgroup
  \def\sline{\noalign{\smallskip\hrule\smallskip}}
  \def\dline{\noalign{\smallskip\hrule\smallskip\hrule\smallskip}}
  \def\\{\cr\sline}
  $$\vbox\bgroup
  \halign\bgroup
    &\strut##\hfil&##\hfil\cr
}
\def\etable{\egroup\egroup$$\egroup}
\begin{table}[tp]
\let\normalsize=\footnotesize\normalsize
\def\eee{\,e\,}
\btable
\dline
  & s.bit (F77)
  & particle (F77)
  & intermed. (C)
  & netw. (C)
  \cr
\dline
Sparc10
  & 4.0 MUPS
  & 3.4 MUPS
  & 1.9 MUPS
  & 1.2 MUPS
  \cr
  & 30\,000 km
  &
  & 14\,000 km
  & 8\,800 km
  \\
PVM
  & 19.0 MUPS
  &
  & 8.9 MUPS
  \cr
($5 \times$ Sp10)
  & 140\,000 km
  &
  & 65\,000 km
  \\
SX-3/11${}^{(1)}$
  & \bf 533 MUPS
  &
  & \bf 2.8 MUPS
  \cr
1 CPN
  & 4\,000\,000 km
  &
  & 21\,000 km
  \\
GCel-3
  & 102 MUPS
  & 211 MUPS
  & 121 MUPS
  \cr
1024 CPNs
  & 750\,000 km
  & 1\,550\,000 km
  & 900\,000 km
  \\
iPSC
  & 83 MUPS
  &
  & 35 MUPS
  \cr
32 CPNs
  & 630\,000 km
  &
  & 260\,000 km
  \\
Paragon
  &
  &
  &
  \cr
64 CPNs
  &
  &
  &
  & \bf 290\,000 km
  \\
CM-5${}^{(1)}$
  & \bf 173 MUPS
  &
  & \bf 30 MUPS
  \cr
32 CPNs
  & 1\,300\,300 km
  &
  & 220\,000 km
  \\
CM-5${}^{(1)}$
  \cr
1024 CPNs
  &
  &
  &
  & \bf [$>$ 1.7\eee 6 km]
  \\
\etable
\noindent ${}^{(1)}$ CPN(s) has/have vector units (SIMD instruction set)

\noindent ${}^{(2)}$ using data parallel Fortran (CMF)

\noindent ${}^{(3)}$ using message passing (CMMD)
\end{table}

}
\caption{\label{cpu}
Computational speed of different algorithms and different machines.
The machines are a SUN Sparc10 Workstation, 5~coupled Workstations
Sparc10 under PVM, a NEC SX-3/11 traditional vectorcomputer, a
Parsytec GCel-3 massively parallel computer with 1024 relatively slow
T805-CPU's, an Intel iPSC/860 Hypercube with 32~nodes, and Intel
Paragon XP-10/S with 64~nodes, and a Thinking Machines CM-5 with
32~nodes, each node containing one Sparc processor and 4~vector
units. --- ``s.-bit'' refers to the single-bit coded site-oriented
algorithm, ``particle'' and ``intermed.''\ to the particle-oriented
and the intermediate one, and ``netw.''\ refers to a network
implementation of the freeway network of Germany. --- All values are
valid for a vehicle density of $c = 0.1$ and system size of $L =
1\,333\,333$ sites, corresponding to 10\,000 single lane kilometers.
MUPS corresponds to Mega-Updates Per Second, i.e., the number of sites
updated per second divided by $10^6$.  The other number is the
extrapolated real time system size, i.e.\ the extrapolated system size
(assuming linear speed-up) where the computation would be just as fast
as reality.  Values in brackets [] are estimates.
}
\end{table}

\bfig
\caption{\label{V1.pixel}
Evolution of different automata rules from an ordered initial state of
density $c = 0.5$. Particles are moving to the right.  {\it
Left:\/} Same as (stochastic) asymmetric exclusion, except that the
update is parallel.  The same plot for exact asymmetric exclusion
looks similar. Note that waves do not move in space. {\it Right:\/}
Same as asymmetric exclusion, except that the maximum velocity is 5.
Waves are now moving backwards, indicating a density above maximum
flow. --- Neither the parallel update nor the higher maximum velocity
alone are sufficient to change the qualitative dynamics of the
asymmetric exclusion model.}
\efig

\bfig
\caption{\label{parallel.pixel}
Evolution of our traffic model (maximum velocity $v_{max}=5$, parallel
update) for density $c=0.3$ (left) and $c=0.1$ (right) from
ordered initial conditions.  Random sequential update with $v_{max}=5$
at density $c=0.3$ (not shown) looks qualitatively similar to
Fig.~\protect\ref{V1.pixel} left, whereas at density $c=0.1$ and
random sequential update the waves are moving to the {\em right} (not
shown).  The density of maximum throughput is much lower for the
parallel update, and instead of the waves switching from backward to
forward motion at this point, they tend to vanish completely (cf.\
\protect\cite{NaP94}).  Moreover, the right picture illustrates the
instability mechanism similar to the continuous model
of~\protect\cite{Kerner.Konh}. }
\efig

\bfig
\caption{\label{fdiag}%
(a) Fundamental diagrams flow $j$ vs.~density~$c$ for maximum
velocities $v_{max} = 1$, 2, $\ldots$, 5, and for a different
fluctuation parameter $p=0.25$ instead of 0.5 at $v_{max}=5$.  (b)
Fluctuations of local speed as a function of density ($v_{max}=5$ and
$p=0.5$).  }
\efig

\bfig
\caption{\label{rho_v}%
Partial densities $c_0(c)$, $c_1(c)$, $\ldots$, $c_5(c)$
(i.e.\ for velocities 0 to 5) for mean field approximation.  $p=0.5$}
\efig

\bfig
\caption{\label{flow.inf}%
Flow for $v_{max}=\infty$ in mean field approximation.  From bottom to
top, the randomization parameter~$p$ is 0.1, 0.3, 0.5, 0.7, and 0.9.
}
\efig

\bfig
\caption{\label{V1}%
Convergence of the $n$-cluster approximations to the simulation result
for the case $v_{max}=1$ and $p=0.5$.  Already the 2-cluster
approximation is exact.}
\efig

\bfig
\caption{\label{V2}%
Convergence of the $n$-cluster approximations to the simulation result
for $v_{max}=2$ and $p=0.5$.  Already the 5-cluster approximation
gives a good fit of the simulation data.}
\efig

}  

\vfill\eject

\end{document}